# Comparing Visual Metaphors with Textual Code For Learning Basic Computer Science Concepts in Virtual Reality

KEVIN BARON, The University of Washington Information School, Seattle, WA, USA

## 1    ABSTRACT

This paper represents a pilot study examining learners who are new to computer science (CS). Subjects are taught to program in one of two virtual reality (VR) applications developed by the researcher that use interactable objects representing programming concepts. The different versions are the basis for two experimental groups. One version of the app uses textual code for the interactable programming objects and the other version uses everyday objects as visual metaphors for the CS concepts the programming objects represent. For the two experimental groups, the study compares the results of self-efficacy surveys and CS knowledge tests taken before and after the VR activity intervention. An attitudinal survey taken after the intervention examines learners' sense of productivity and engagement with the VR activity. While further iterations of the study with a larger sample size would be needed to confirm any results, preliminary findings from the pilot study suggest that both methods of teaching basic programming concepts in VR can lead to increased levels of self-efficacy and knowledge regarding CS, and can contribute toward productive mental states.

## 2    KEYWORDS

virtual reality, education, computer science, programming, visual learning, textual learning

## 3    INTRODUCTION

Prior research in CS education indicates that multiple approaches have been used to reduce barriers to entry to the field. This has included traditional forms of CS education and newer forms of VR CS education.  Consideration of existing approaches has led the current study to focus on teaching basic CS topics and comparing the use of text-based interactable programming objects with the use of visual metaphors consisting of everyday objects used as representations of programming concepts.

To compare these two versions of teaching programming concepts, the researcher used the 3D application development tool Unity to design, develop, and test a customized VR application for stand-alone usage on Meta Quest 2 headsets. The experience within the app guides subjects through miniature lessons and allows them to interact with programming objects to create their own programs. Two different versions of the app were produced, one for each experimental condition to be used as an intervention activity.

The researcher also developed questionnaires and knowledge tests for subjects to take before and after the VR activity intervention to gauge effectiveness of each of the two versions of the app. Questionnaire and test responses were analyzed to produce the results of the study.

## 3.1  2D COMPUTER SCIENCE EDUCATION

Because computer science is a field with a high level of perceived difficulty and a high dropout rate for students in introductory courses, there have been efforts made by multiple stakeholders to reduce the barriers to entry by teaching beginner programming concepts [1].

Some of these efforts have taken the form of free and paid online courses that teach a particular programming language or a selection of programming languages through a curriculum consisting of progressive exercises beginning with fundamental concepts and easing into more advanced topics. Examples of such options include Codecademy, Udacity, Coursera, The Odin Project, Codewars, FreeCodeCamp, Udemy, edX, Code.org, as well as others [2]. Schools like UC Berkeley are among educational institutions that offer their own coding bootcamp style programs, as they're also known [3].

While these options are marketed as being ideal for novices of most age groups, they commonly use the pedagogical format of typing textual code into a window. This format is similar to real programming in a production context, but it may be better suited for older learners such as college students and continuing professionals who already have relatively high levels of reading and writing comprehension. The format may not be as accessible and engaging for younger learners as more visual programming options, especially learners young enough not to have developed high-level reading and writing skills [4] [5].

The need for more visually based educational programming options specifically suited for younger learners led to the creation and expansion of block-based programming (BBP) tools such as Blockly, Scratch, CoBlox, and others [4]. These tools create a graphical user interface (GUI) for the user to be able to manipulate and combine pieces of code to create meaningful animations and interactions [4]. This format has demonstrated success in the areas of engagement and learning new programming concepts when compared with text-based approaches [5].

A 1999 study conducted by Mckay explored the effects of graphical metaphors on learning CS in adults [6]. The study uses the term "graphical metaphor" synonymously with the current study's term "visual metaphor," meaning a non-textual image which abstractly represents a concept [6]. The only difference in usage is that Mckay's study involved 2D images, whereas the current study uses 3D objects in the VR activity [6]. To simplify the terminology used, "visual metaphor" will be used in place of "graphical metaphor" for further explanation of Mckay's study.

Mckay demonstrated that visual metaphors can potentially improve learning performance for beginning CS concepts in adults [6]. However, the most common BBP options simply use the visual nature of the block structure and stop short of using visual metaphors for the programming concepts [5]. Instead, they put small bits of text into the blocks to represent their functionality.

The use of visual representations of HTML tags in the form of emojis by Codemoji is one use case of visual metaphor for coding concepts, but this example is language specific and does not encapsulate the general programming concepts such as data types and sequences [7].

## 3.2 VIRTUAL REALITY IN COMPUTER SCIENCE EDUCATION

In recent years, there has been increasing availability of VR headsets with features like six degrees of freedom, independent usage without needing to be tethered to a PC, and internal position tracking. The increased availability has opened new possibilities for learning interactivity and engagement in many subjects.

For example, a study tracked and surveyed undergraduate physics students learning in VR at the University of Washington (UW) [8]. The study demonstrated that students were able to engage as productively in VR lab sessions as in traditional lab sessions [8]. The VR lab allowed for the creation and observation of novel physics phenomena that were not possible in real settings. This gave students the opportunity to create and test against their own theories for the contrived laws of physics in play [8]. VR has been used in education for a variety of other subjects as well, and has shown that it can widely improve immersion and engagement for most subjects, particularly for those involving places that are abstract or difficult to access [9] [10].

Computer science concepts are abstract, and because of this, can be difficult to access in a physical way other than at a computer on a programming console or in a simulation of one like coding bootcamps commonly use. This is where the field presents the opportunity to particularly benefit from VR education techniques.

In a 2021 literature review of VR in CS education, Agbo et al. indicate that there was a rise in scholarly articles for this subject starting in 2017 and continuing through that paper's publication [11]. This coincides with the continuing release of less expensive and more usable VR systems. The authors call for more efforts to be made by stakeholders across countries and institutions, and specifically for more rigorous methodological approaches in future studies in order to produce more evidence-based research output [11]. These considerations have been taken into account in designing the methodology for the current study.

While VR's immersive and engagement capabilities could be used to further explore the opportunity of visual metaphors for CS concepts from beginning concepts to more advanced ones, there is little research into the effectiveness of visual metaphors for beginning concepts. There have been, on the other hand, examinations of its effectiveness for teaching object-oriented programming (OOP) in particular [12] [13]. These studies indicate that completing interactive activities using visual metaphors and analogies, such as the parts of a house for the different sub-areas of object-oriented programming, can be effective at improving comprehension and confidence regarding these sub-areas. This was found to be especially true for sub-areas associated with particularly low levels of comprehension and confidence before exposure to the visual metaphors in VR, which were targeted in the application design [12]. It is noted in one study that similar gains in comprehension and confidence were achieved with traditional text-based methods [13].

The state of prior work has led the current research toward a study which empirically examines the effectiveness of VR in teaching beginning programming concepts. It is intended that the study will positively contribute toward this specific area where there is the largest lack of similar research compared with other more advanced CS topics. The study will be guided by the following research question. For learners who are new to computer programming, how do VR learning experiences with interactable programming objects compare in effectiveness when textual code is used for the programming objects versus when visual metaphors are used for the programming objects?

Measures of effectiveness include change in basic CS knowledge, change in self-efficacy related to basic programming concepts, and productive engagement during the learning activity.

## 4   METHODS

## 4.1  VIRTUAL REALITY APPLICATION INTERVENTION

The study's six subjects were split evenly into two experimental groups, A and B. To isolate the effectiveness of visual metaphor in teaching basic CS concepts compared to more text-based programming, the study controlled the pedagogical medium by using VR activity interventions for both experimental groups in between a pre-intervention questionnaire and knowledge test, and a post-intervention questionnaire and knowledge test.

Group A uses a version of the app that employs interactive rectangular text-based objects as physical representations of functional pieces of code and raw data, such as strings, Booleans, and integers (see Figures 1-5). Group B uses a version of the app that employs interactive visual metaphors for pieces of code and data. The visual metaphors are 3D models of everyday objects that abstractly represent the functional pieces of code and raw data (see Figures 1-5). For example, version A uses a text box with the phrase "Print:" and version B uses a lifesize model of a desktop printer. The text boxes and other models are scaled to be approximately the same size.  See "Appendix A: Computer Science Concept Representations in Experimental Groups" for a comparison of text used in experimental group A and visual metaphors used in experimental group B. Besides these differences and the associated minor wording differences in the in-app instructions subjects followed, the learning exercises were identical and had the following features. These features were designed, implemented, and tested by the primary researcher in the weeks before the study sessions were conducted.

In the app, the subject is standing in the middle of an open area with a metallic floor and views of outer space. This is part of the theme tying together the various tasks within the activity where the subject is explained to be the lawyer of an astronaut in a trial case.

The subject can see their VR hand controllers represented as white spheres in the virtual space. The controller spheres have lasers projecting out of the front, showing where the subject is pointing with either hand. The lasers can be used to click buttons, grab objects, and teleport to different locations on the floor as needed.

To the subject's left is a panel showing what level the subject is currently on (see Figure 6). The panel has buttons that can be clicked to navigate between the tutorial and levels 1-4. The panel indicates which of these levels are complete and incomplete at any given moment. To the subject's right is a panel with instructions for each level and buttons to navigate those instructions (see Figure 6).

Also to the subject's left is a print log panel that shows outputs printed by the subject as they create their programs (see Figure 6). All tasks involve printing specific messages to the print log, and whether those messages have been printed for each level determines if that level is complete. Above the print log panel is a "Run Program" button that can be clicked to run a program created by the subject.

Directly in front of the subject are the available programming objects for a given level. These can be used to create programs by connecting them together. Some programming objects have green transparent zones attached to them that can be used to connect programming objects in sequence. Some programming objects have blue transparent zones that represent their input parameters used to perform their functions such as printing or adding two integers. These blue transparent zones can be filled with different programming objects to produce different results when the program runs.

The tutorial covers basic orientation for the subject as far as the theme of the tasks. It also explains how to navigate the instructions and the level display panel, move around and manipulate programming objects, connect programming objects properly to create a program, run the program, verify that the output has been printed to the print log, and verify that the tutorial is complete.

Level one covers data types and sequences, and explains the process of printing to the print log in more detail than is covered in the tutorial. Level two covers arithmetic operations, string concatenation, and nesting inputs. Level three covers Boolean values True and False, and how to produce them using comparison operators like equality checking (==) and greater than (>), as well as logical operators AND (`&&`), OR (`||`), and NOT (`!`). Level four covers conditional statements (`if`) and touches on variables.

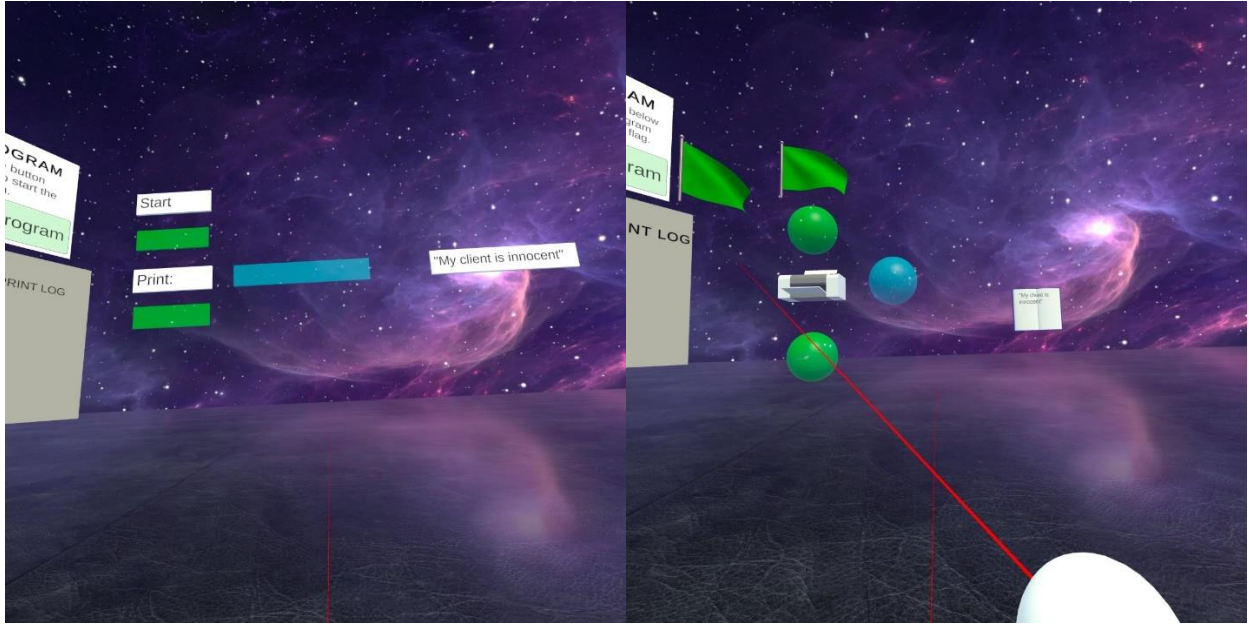

Figure 1. Tutorial programming materials. Left: text box format used for group A. Right: visual metaphors used for group B. In the group B image, the user's right controller sphere is visible at the bottom of the image with the laser pointer pointing toward the printer.

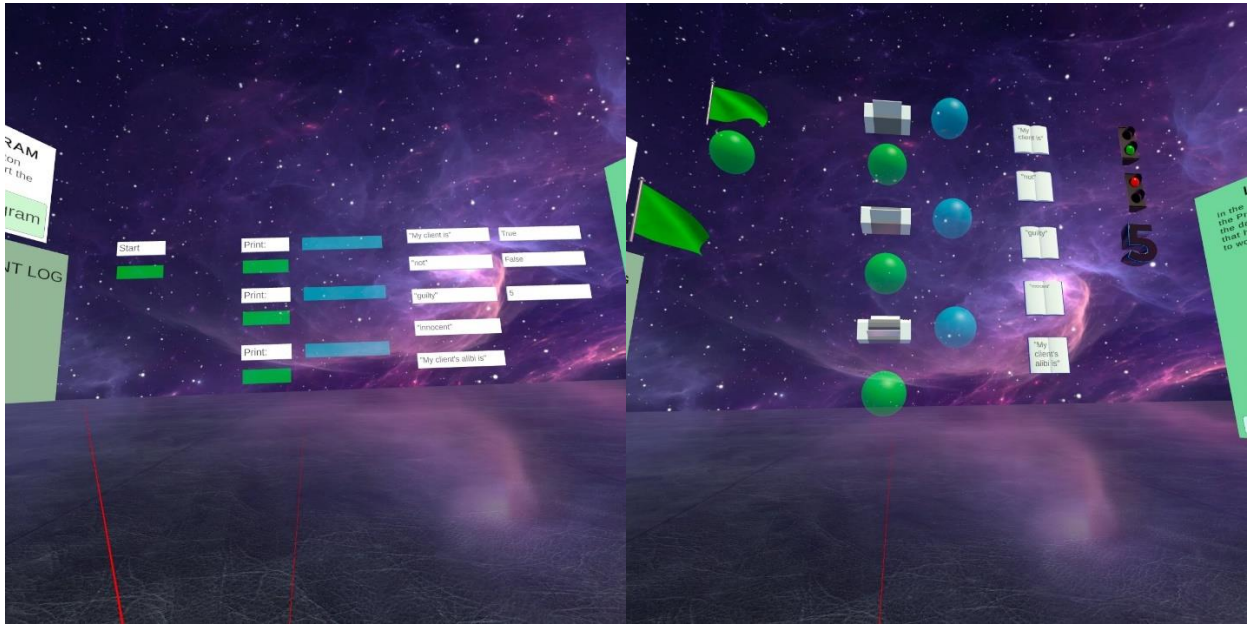

Figure 2. Level 1 programming materials. Left: text box format used for group A. Right: visual metaphors used for group B.

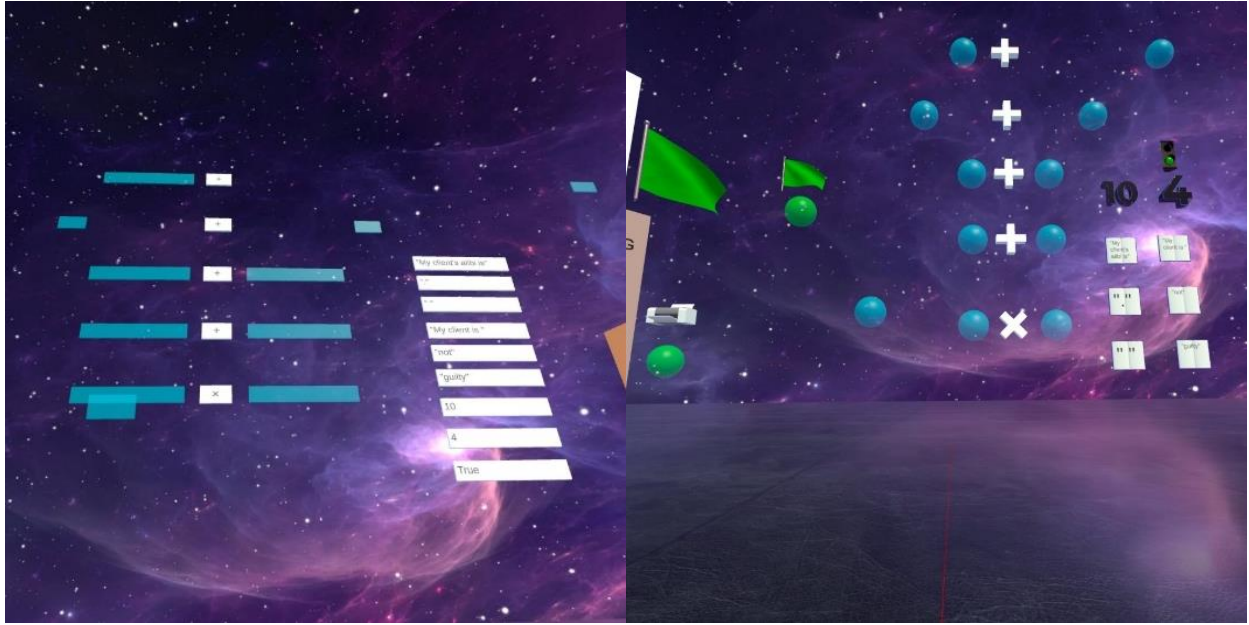

Figure 3. Level 2 programming materials. Left: text box format used for group A. Right: visual metaphors used for group B.

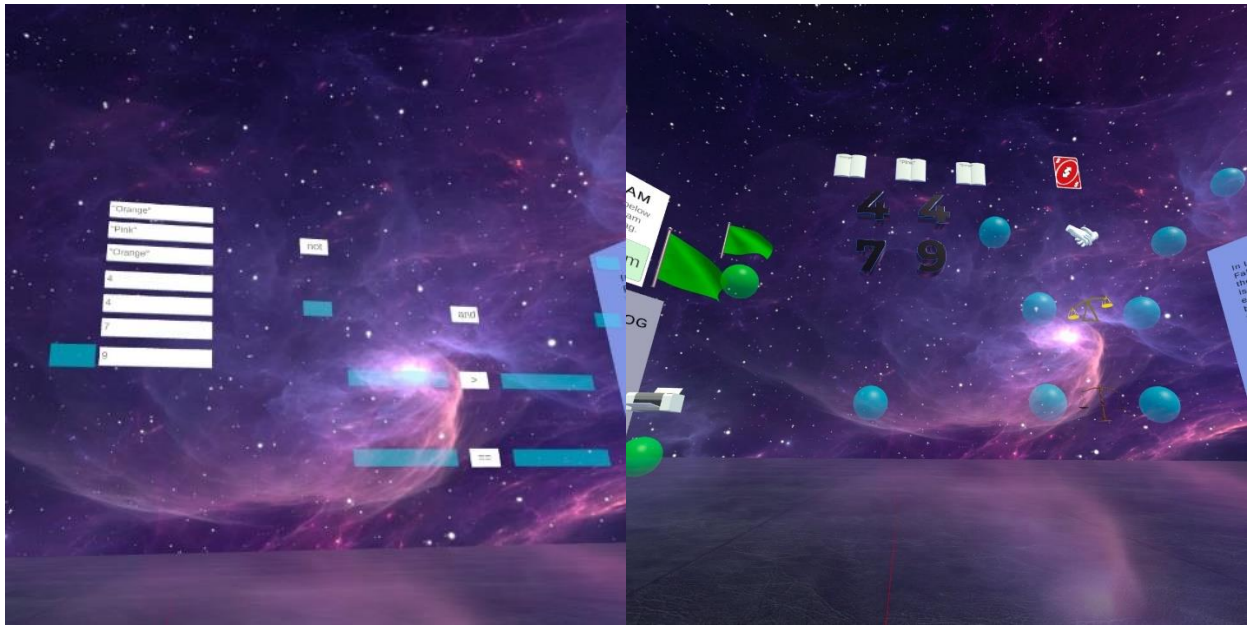

Figure 4. Level 3 programming materials. Left: text box format used for group A. Right: visual metaphors used for group B.

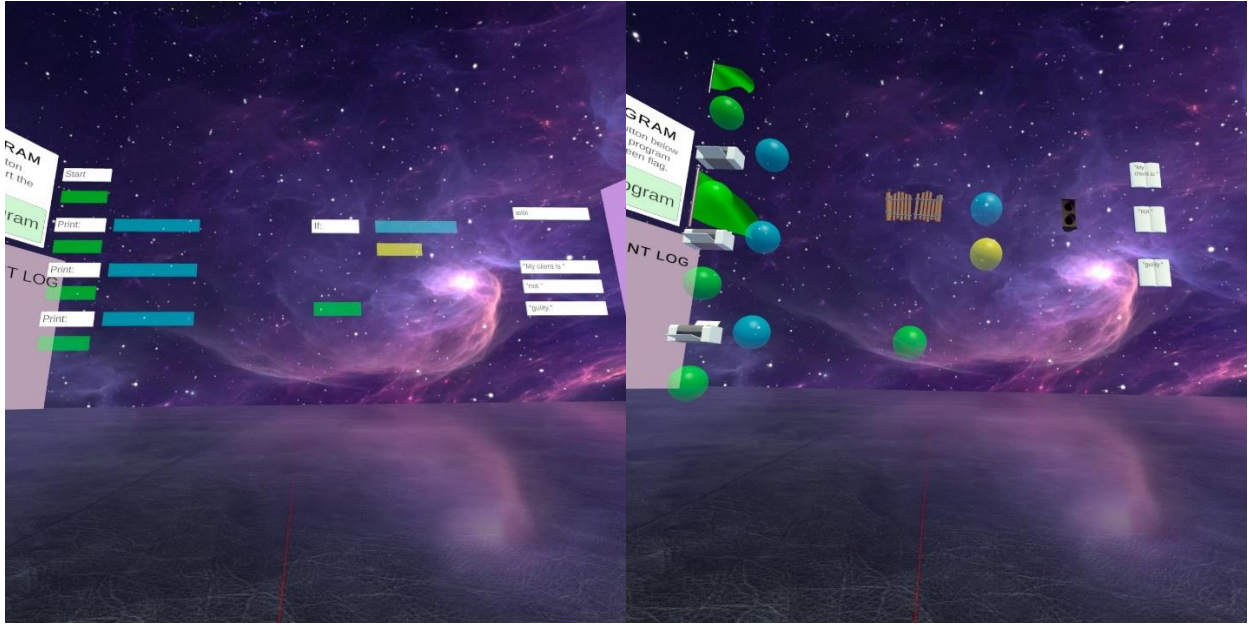

Figure 5. Level 4 programming materials. Left: text box format used for group A. Right: visual metaphors used for group B.

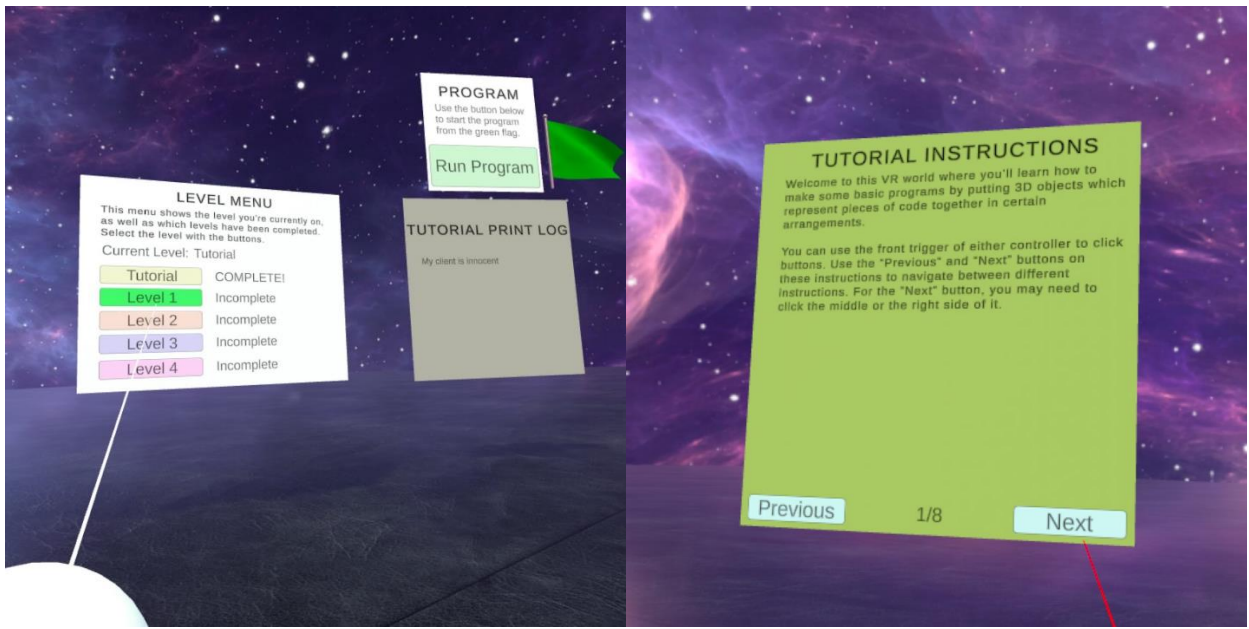

Figure 6. Functional User Interfaces. Left: Panels at subject's left in VR app. These are the Level Menu, the "Run Program" button, and Print Log with message printed. Printing this message causes the status next to the "Tutorial" button on the Level Menu to indicate "COMPLETE!" The left controller laser pointer is hovering on Level 1 button. Right: Tutorial Instructions panel at subject's right in VR app.

## 4.2  PRE- AND POST-INTERVENTION QUESTIONNAIRES AND TESTS

The study uses an identical pre-intervention questionnaire and post-intervention questionnaire to measure subjects' change in self-efficacy regarding basic CS concepts and reasoning strategies. The five questions in the questionnaire gauge confidence in using basic CS concepts and reasoning strategies on a scale of 1 to 7, with 1 representing "not at all confident" and 7 representing "completely confident." The questions were modeled off of similar questions used in Canright's and Brahmia's UW VR physics lab study, and are rooted in the philosophy of self-efficacy being part of a unified theory for behavioral change [8] [14]. See "Appendix B: Questionnaires and Tests" for the questions used.

The study also uses pre-intervention and post-intervention knowledge tests to determine the change in knowledge of basic CS concepts including data types, print statements, arithmetic and comparison, string concatenation, sequencing, nesting, and conditionals. The assessment of these subject areas was guided by their identification as key areas for new programmers [15]. Knowledge questions were adapted from the AP Computer Science A Exam and abstracted to programming concepts so that they could be answered after completing either version of the VR activity [16]. Other studies aimed at developing assessment tools for CS knowledge using the abstraction of block-based programming also helped to guide and validate the construction and adaptation of the questions used for the current study [17] [1].

The pre-intervention test and post-intervention test both consist of 10 questions each, with eight questions being multiple choice, and two questions being multiple option selection where more than one answer response could be chosen for a question. Five questions are identical between the two tests and five questions are changed slightly from the pre-intervention test to the post-intervention test while keeping the fundamental concept tested for the same. Changes pertain to numbers used in the evaluation, operators being switched between equality and the greater than operator or between AND (`&&`) and OR (`||`) operators. See "Appendix B: Questionnaires and Tests" for the questions used in the pre-intervention test and the post-intervention test.

The post-intervention questionnaire also includes seven questions evaluating the subjects' engagement. Each question associates different meanings to a 1-7 Likert scale. The questions were modeled off similar questions used in Canright's and Brahmia's UW VR physics lab study, which themselves were based on a method of associating responses to a flow state scale used to measure optimal user experience [8] [18]. See "Appendix B: Questionnaires and Tests" for the questions used.

## 4.3  STUDY SESSIONS

During the two-hour study sessions, all subjects were blind to the fact that there were multiple versions of the VR activity. Once they had arrived to the study session room, subjects were asked if they had reviewed the consent form emailed out prior and if they had any questions regarding it. Once any questions were answered and it was made clear that any further questions about completing the study could be asked at any time, the subjects started the pre-intervention questionnaire and test.

Once subjects had completed the pre-intervention questionnaire and test, they were shown how to fit the Meta Quest 2 devices onto their heads and how to make sure that it was comfortable and that they could see and read clearly. Because some subjects had prior experience with VR headsets and some had no experience, basic controls available on the controllers were reviewed by the researcher at the transition into the headset. These controls are also covered in the tutorial within the VR activity.

Once the subjects started the VR activity, the researcher was able to view what the subject was seeing in VR through a screen cast to a computer to more easily answer any questions the subjects had along the way. For the first subject, the screen cast process had not been set up yet, but the researcher was able to follow along verbally with the subject for any questions asked.

Once the subjects finished the VR activity, they were offered to take a break if they wanted to, and then moved on to the post-intervention questionnaire and test. Once they were done with the post-intervention questionnaire and test, the study session was complete.

## 4.4  SUBJECT RECRUITMENT

Because this research study was conducted as an Informatics Capstone at the UW, its timeline was condensed into the ten weeks of the Winter 2024 quarter. The compressed nature of the timeline made it logical to apply for exempt research status through the UW Institutional Review Board (IRB). One requirement for exempt status was for all subjects to be age 18 or older.

A screening survey form was sent out via UW departmental listservs to students and faculty in various departments. It was also sent out on UW-affiliated interest-based online chat servers.

One question on the screening survey screened for age of participants. Two questions assessed participant eligibility for the study pertaining to recency of experience with CS or programming. One question asked about curricular programming experience and the other asked about extracurricular programming experience. Respondents who initially self-reported as having had either form of experience with programming within the past two years were screened out from participating in the study.

The primary researcher manually screened responses and performed outreach to willing participants. Between the eligibility criteria, follow-up communication regarding study session logistics, and availability of participants during the primary window used for study sessions, seven subjects participated in the study. Of those seven, the results for six subjects were able to be used for the study. All of the six subjects whose results were used had initially responded that they had never had any experience with computer programming.

The two screening questions assessing recency of computer programming experience were asked again to subjects on the pre-intervention questionnaire at the time of a given study session. At this time, two subjects responded that they had had experience within the past two years. Because of the discrepancy between initial screening responses and responses at the time of the study session, these two subjects were asked about the extent of the recent experience, and their responses made it clear that the recent experience could be considered negligible within the context of the study and that their results would still be usable.

The one participant who had initially reported experience in computer programming more than two years before the time of the initial screening survey also changed the response at the time of the study session to having had experience within the past two years. Because a follow-up response was not received from this subject as to the extent of this experience, it could not be determined that it was negligible within the context of the study, so that subject's data was not included in the final results.

## 5   RESULTS

## 5.1 COMPUTER SCIENCE KNOWLEDGE

While the average post-intervention CS knowledge test score of subjects in experimental group B, who had used visual metaphors in their VR activity, was slightly higher than the average post-intervention test score of subjects in experimental group A (Group A average: 6.7 (n=3); Group B average: 7 (n=3)), group B had also shown higher pre-intervention test scores (Group A average: 1.7 (n=3); Group B average: 3.7 (n=3)).

This resulted in Group A demonstrating higher levels of positive change in test score between pre-intervention test and post-intervention test (Group A average change: +5 (n=3); Group B average change: +3.3 (n=3)).

## 5.2  COMPUTER SCIENCE SELF-EFFICACY

For changes in self-efficacy between pre-intervention and post-intervention, when averaged together, subjects from group B had greater improvements in self-efficacy, or confidence, for four out of the five questions answered. Between the groups, subjects in group B had consistently lower confidence ratings in the pre-intervention questionnaire than subjects from group A.

Subjects from group A had consistently higher average confidence ratings in the post-intervention questionnaire. However, because subjects had rated their confidence as higher in the pre-intervention questionnaire, the shifts were not as large in group A as they were in group B, except for question 1. Question 1 has to do with identifying basic data types used in programming. For this question, group A demonstrated higher pre-intervention and post-intervention average confidence ratings as well as a larger shift in confidence rating from pre-intervention to post-intervention.

## 5.3  PRODUCTIVE ENGAGEMENT IN VR ACTIVITY

The psychological theory of flow pioneered by Csíkszentmihályi is used to examine subjects' engagement with the VR activities [19]. Flow is described as a mental state in which one is completely absorbed in an activity for its own sake, where one action leads smoothly into the next, and one's sense of time can become distorted [19].

Determination of mental state is made by associating the subject's self-perceived knowledge and skillfulness at an activity with the X-axis of a diagram called a flowplot. The subject's self-perceived challenge posed by the activity is associated with the Y-axis of the flowplot and the subject is plotted on the flowplot at a point with the resulting (X, Y) coordinate.

Flowplots are divided into eight sections representing mental states that correspond with different combinations of low, medium or high perceived skill and knowledge, and low, medium, or high perceived challenge (see Table 1). Also see Figures 8 and 9, which show how the flowplot is divided into these mental states.

Table 1. Flowplot Mental States

| Mental State | Perceived Skill/Knowledge Level | Perceived Challenge Level |
|---|---|---|
| Flow | high | high |
| Arousal | medium | high |
| Control | high | medium |
| Relaxation | high | low |
| Boredom | medium | low |
| Apathy | low | low |
| Worry | low | medium |
| Anxiety | low | high |

The flow state is where the most effective learning is possible and is most ideal [20]. The flow state, the arousal state and the control state are all considered states of productive engagement [20]. All other states are considered unproductive in terms of engagement [20].

Using questions 5 and 6 from the productive engagement (PE) section of the post-intervention questionnaire (which correspond with questions 6 and 7, respectively, on the entire post-intervention questionnaire), Karelina et al.'s analysis methods are followed to create flowplots [21]. PE question 6 is used as a measure of perceived knowledge and skill on the X-axis, and PE question 5 is used as a measure of challenge on the Y-axis.

As can be seen in Figures 8 and 9, only one subject was in the flow state while participating in the VR activity, and this subject was in group B, using visual metaphor programming. Both experimental groups had one subject in a productive state of engagement [21]. Both groups also had one subject in the relaxation state, indicating that while they felt knowledgeable and skillful, that the presented challenge was relatively low. Only group B had a subject in the anxiety state.

For other questions gauging productive engagement on a 1-7 Likert scale, group A reported on average to have needed quite a bit more assistance from the researcher. Where the scale for the question regarding the level of assistance needed corresponded with 1: Not at all, and 7: A lot, group A reported an average of 5.7, and group B reported an average of 3.3. Group A's lowest individual response value of 4 was nearly the same as group B's highest individual response value response of 5.

Group B reported a slightly higher average value for knowing what to do, as far as the goal of the tasks (5.7) compared to group A (5). Similarly, group B reported a slightly higher average value for knowing how to accomplish the goal of the tasks (5.3) compared to group A (4.7). Group B also reported a slightly higher average value for knowing how well they were doing (5.3) compared to group A (4.7).

Both groups reported an average value of 5.3 for how fun and interesting the VR activity was, where the scale corresponded with 1: Not at all, and 7: Extremely. Group A had shown a slightly wider range of responses to this question, with a set of [4, 5, 7], compared to the set of group B's responses [4, 6, 6].

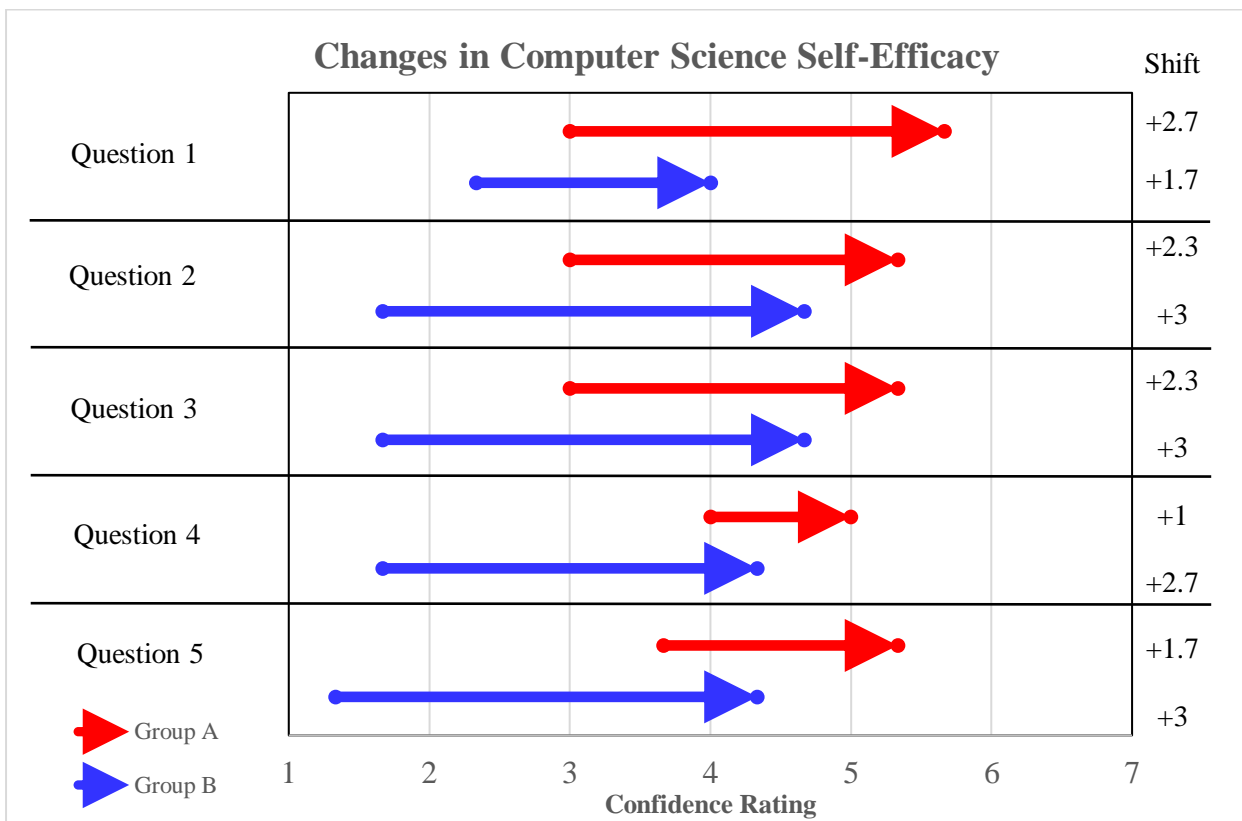

Figure 7. Changes in Computer Science Self-Efficacy. Group A is represented in red, and group B is represented in blue. Question 1 for CS self-efficacy corresponds with question 4 on the pre-intervention questionnaire and question 9 on the post-intervention questionnaire. The rest of the questions correspond sequentially. The tail end of each arrow represents the average confidence rating from subjects on the pre-intervention questionnaire, and the head end represents the average confidence rating on the post-

intervention questionnaire. Changes from pre-intervention to post-intervention are represented by the magnitude of the arrow and are given numerically at the right side of the figure as the "Shift" value.

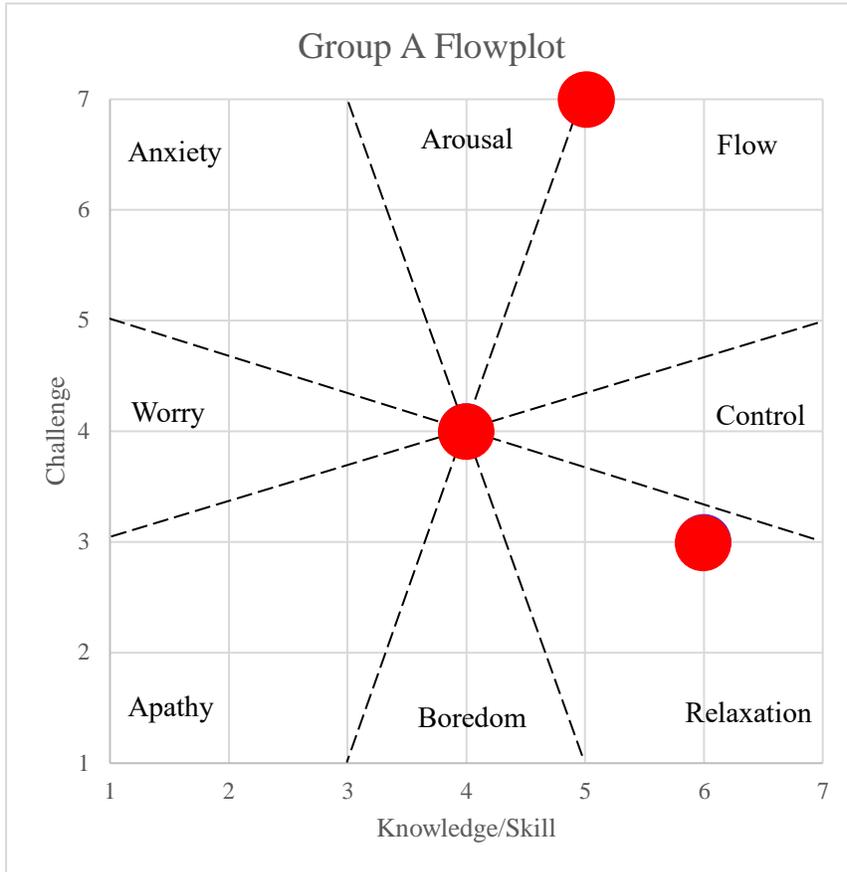

Figure 8. Group A Flowplot. The three red dots represent the three subjects in group A. One subject was on the border of arousal and flow state, one subject was completely neutral between all states, and one was in the relaxation state.

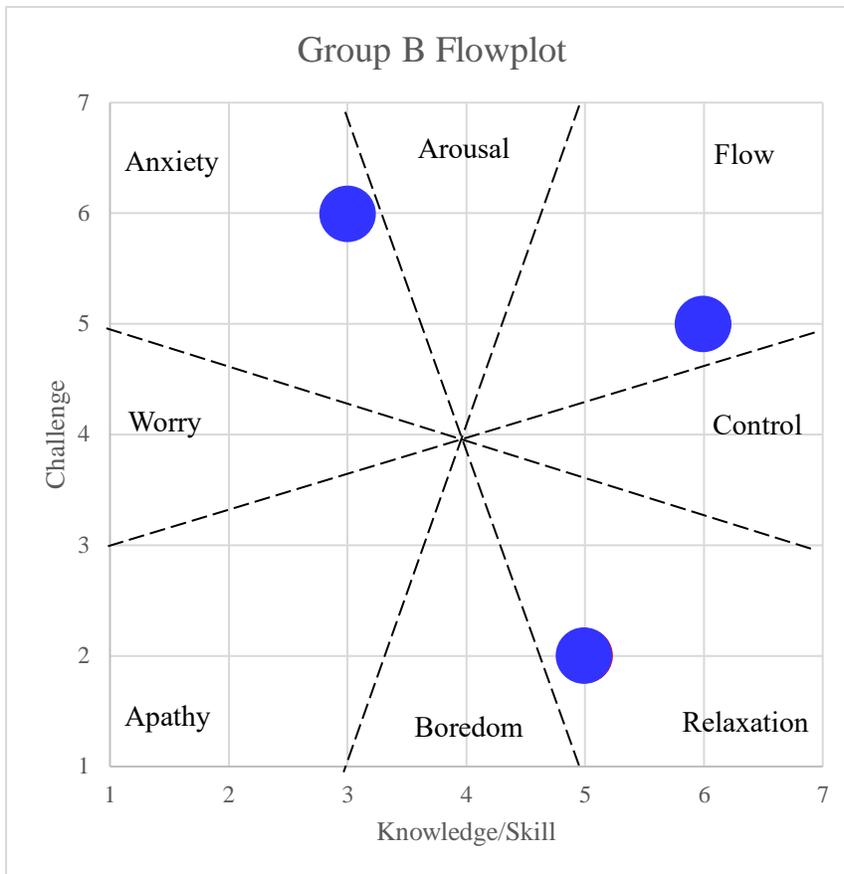

Figure 9. Group B Flowplot. The three blue dots represent the three subjects in group B. One subject was in the flow state, one subject was in the anxiety state, and one was in the relaxation state.

## 6    DISCUSSION

It is interesting to note that while it was theoretically possible to be able to answer all of the questions on the post-intervention test correctly after having done either VR activity, the maximum score achieved on the post-intervention test was 8/10. The two questions missed by the one subject with that score were the two most commonly missed questions by other subjects as well, one regarding space characters and in string concatenation and the other regarding the different types of logical operators.

This difficulty in these questions could potentially be attributed to the minimal instructional coverage of their material and opportunity for practice in the VR activity. Because it was a concern that subjects might not be able to complete all the tasks and the questionnaires and tests in their allotted times, abbreviations of the lesson material had been made in the VR activity design.

While it was noted to pay close attention to spaces while concatenating strings, it was not explained why the subject should pay close attention, nor was there further explanation about some of the intricacies of string concatenation. It was expected that subjects might learn by trial and error, which some did, but others seemed to be able to get through the string concatenation with spaces task without fully grasping the nuance of the task.

Similarly, for logical operators, while the existence of the OR (| |) operator was explained in the instructions, it was not materialized as a programming object that could be interacted with by subjects. This could partially explain why subjects were not able to remember about that operator by the time they came to the post-intervention question about how many basic logical operators there were. The AND (&&) and NOT (!) operators were both available to engage with in the VR activity.

For changes in self-efficacy, the fact that subjects from group B self-reported consistently lower average confidence ratings regarding CS concepts in the pre-intervention questionnaire compared to subjects from group A could have been a driving factor in their ability to report higher confidence ratings in the post-intervention questionnaire, whereas subjects from group A may not have had the same margin for growth in their own determinations of confidence ratings. Larger sample sizes for both groups would ideally reduce the difference between pre-intervention confidence ratings by shifting both groups toward a mean value.

As for the mental states of the subjects while engaging in the VR activities, it is interesting that group B had a wider spread of response values, while group A had more instances of subjects on the border between two or more states (all eight states in one case). It would be even more interesting to see how the flowplots would fill up with more subjects to see if this is an emerging pattern or if it is created by relative outliers. Group A had a narrower spread of responses on the two-dimensional flowplot. The only response of the maximum level of challenge of 7 was found in group A.

It's possible that the higher incidence of perceived assistance required in group A was related to higher occurrences of app crashes for subjects in this group, or to the extra effort involved in snapping the flatter text boxes into the transparent input and execution sequence slots of other text box programming objects. The text boxes had thinner colliders (autosnapping zones which help snap pieces into place) compared with the transparent sphere colliders in the B version of the app.

The larger colliders in the B version were deeper on the relative Z-axis and therefore may have been easier to target while manipulating programming objects to build portions of code.

One subject experienced VR motion sickness throughout the study, and took multiple breaks, which was encouraged, but was still able to complete the VR activity. No other subjects reported feeling motion sickness when asked during or after the VR activity.

Because in the compressed research timeline, not all bugs in the VR app were able to be remedied before the study sessions began, it was known that there was a possibility of the app crashing for both versions. A crash could occur if certain programming objects that should not be connected together were connected, which would cause a memory leak and the VR device would run out of available memory.

The possibility of a crash was explained verbally to participants as they entered the VR activity, and was also explained in the instructions for the levels where crashes were possible. It was explained that if this were to happen, the researcher would restart the app and set the subject to resume where the app had crashed. Preparing subjects for this possibility mitigated the disruption to the learning experience as it was perceived by subjects. Some subjects in both experimental groups completed all levels of the VR activity without any app crashes.

In both pre-intervention and post-intervention knowledge tests, there was an answer option of "I don't know." This led to some scores, particularly in the pre-intervention test, being lower than they might otherwise have been by being forced to guess at answers. With enough subjects, this option could actually benefit the results by eliminating the correct guess answers. But with such a small sample size, an answering pattern of mostly "I don't know" answers being used by one or a few subjects may have heavily swayed the results, depending on which experimental groups those subjects happened to be a part of.

## 8   CONCLUSION

While further iterations of the study with a larger sample size would be needed to confirm any results or generalize conclusions to the population sampled from, preliminary findings from the pilot study suggest that both methods of teaching basic programming concepts in VR can lead to increased levels of CS self-efficacy and CS knowledge and can contribute toward productive mental states.

In comparing the results between the two experimental groups, it is possible that programming with visual metaphors contributes to greater improvements in CS self-efficacy than programming with text-based objects. It is also possible that programming with visual metaphors is more likely than programming with text-based objects to induce the most ideal mental state for productive engagement, the flow state.

Another potential conclusion is that programming with visual metaphors can result in higher overall levels of CS knowledge than programming with text-based objects, but that programming with text-based objects can contribute to greater improvement in CS knowledge when knowledge levels prior to intervention are taken into account.

## 9    FURTHER RESEARCH

Future iterations of this pilot study with enough available time could conduct a long-form version of the study consisting of multiple study sessions per subject and covering a wider range of beginner and potentially even intermediate CS topics.

Future iterations might also consider conducting the experiment with younger subjects, as programming with non-textual interfaces and environments could be better suited to younger populations with less reading and writing experience [15].

Further variations of this study could also benefit by using additional experimental groups that more closely model traditional 2D styles of teaching found to be common across various coding boot camps. These additional experimental groups could provide a baseline to compare the results of the two VR-based experimental groups [2]. This could include a text-based coding miniature curriculum that is most similar to the traditional coding course offerings. It could also include a version of the same curriculum that uses 2D programming object analogs of the 3D visual metaphors used in the current VR version of the study. Comparing subjects' engagement and changes in knowledge and self-efficacy for one or both of these 2D versions of the curriculum to the VR versions could show how effective VR is as a tool for teaching CS concepts with respect to baseline teaching tools.

## ACKNOWLEDGMENTS

I would like to thank my research adviser John Akers and my Informatics Research Capstone instructor of record Dr. Katie Davis both for supporting this research. I would also like to thank the UW Reality Lab, of which John is the Director, for providing a space to meet and conduct study sessions, for access to the VR headsets used in the study, and for the opportunity to collaborate with its Thunder Struct team as part of the lab's Incubator. Thanks also to Informatics Advisor Elisa Tran for broadcasting the message about recruiting for the study to advisers of other UW departments.

## REFERENCES

[1]   Andrew Luxton-Reilly, Brett A. Becker, Yingjun Cao, Roger McDermott, Claudio Mirolo, Andreas Mühling, ... & Jacqueline Whalley. 2018. Developing assessments to determine mastery of programming fundamentals. In Proceedings of the 2017 ITiCSE Conference on Working Group Reports (ITiCSE-WGR '17), 47-69.

[2]   Scott Morris. 2024. 100+ Free Online Websites To Learn To Code For Beginners. https://skillcrush.com/blog/64-online-resources-to-learn-to-code-for-free/#1

[3]   Anon. 2024. Berkeley coding boot camp: Online: San Francisco & Bay Area. February 2024. Retrieved March 9, 2024 from https://bootcamp.berkeley.edu/coding/

[4]   David Weintrop. 2019. Block-based programming in computer science education. Commun. ACM 62, 8 (August 2019), 22–25. https://doi.org/10.1145/3341221

[5]   David Weintrop and Uri Wilensky. 2017. Comparing Block-Based and Text-Based Programming in High School Computer Science Classrooms. ACM Trans. Comput. Educ. 18, 1, Article 3 (March 2018), 25 pages. https://doi.org/10.1145/3089799

[6]   Elspeth Mckay. 1999. Exploring the effect of graphical metaphors on the performance of learning computer programming concepts in adult learners: a pilot study. Educational Psychology, 19(4), 471-487.

[7]   Anon. 2024. Codemoji Playground. Retrieved March 9, 2024 from https://www.codemoji.com/playground.php

[8]   Jared P. Canright & Suzanne White Brahmia. 2023. Modeling novel physics in virtual reality labs: An affective analysis of student learning. arXiv preprint arXiv:2310.07952.

[9]   Andreas Marougkas, Christos Troussas, Akrivi Krouska, & Cleo Sgouropoulou. 2023. Virtual reality in education: a review of learning theories, approaches and methodologies for the last decade. Electronics, 12(13), 2832.

[10]  David Hamilton, Jim McKechnie, Edward Edgerton, and Claire Wilson. 2021. Immersive virtual reality as a pedagogical tool in education: a systematic literature review of quantitative learning outcomes and experimental design. Journal of Computers in Education, 8(1), 1-32.

[11]  Friday Joseph Agbo, Ismaila Temitayo Sanusi, Solomon Sunday Oyelere, and Jarkko Suhonen. 2021. Application of virtual reality in computer science education: A systemic review based on bibliometric and content analysis methods. Education Sciences, 11(3), 142.

[12]  Tevita Tanielu, Raymond 'Akau'ola, Elliot Varoy, and Nasser Giacaman. 2019. Combining Analogies and Virtual Reality for Active and Visual Object-Oriented Programming. In Proceedings of the ACM

Conference on Global Computing Education (CompEd '19). Association for Computing Machinery, New York, NY, USA, 92–98. https://doi.org/10.1145/3300115.3309513

[13] Victor Lian, Elliot Varoy, and Nasser Giacaman 2022. Learning Object-Oriented Programming Concepts Through Visual Analogies. IEEE Transactions on Learning Technologies, 15(1), 78-92.

[14] Albert Bandura. 1977. Self-efficacy: Toward a unifying theory of behavioral change. Psychological Review 84, 191.

[15] Karen Brennan and Mitchel Resnick. 2012. New frameworks for studying and assessing the development of computational thinking. In Proceedings of the 2012 annual meeting of the American educational research association, Vancouver, Canada (Vol. 1, p. 25).

[16] Anon. 2024. AP Computer Science A Exam. Multiple-Choice Practice. AP Computer Science A Exam. Retrieved February 1, 2024, from https://www.apcsaexam.org/mcpractice.html

[17] Arif Rachmatullah, Bita Akram, Danielle Boulden, Bradford Mott, Kristy Boyer, James Lester, and Eric Wiebe 2020. Development and validation of the middle grades computer science concept inventory (MG-CSCI) assessment. EURASIA Journal of Mathematics, Science and Technology Education, 16(5), em1841.

[18] Susan A. Jackson and Herbert W. Marsh. 1996. Development and Validation of a Scale to Measure Optimal Experience: The Flow State Scale. Journal of Sport and Exercise Psychology 18, 17.

[19] Mihaly Csíkszentmihályi. 1990. Flow: The Psychology of Optimal Experience. Harper & Row.

[20] Mihaly Csíkszentmihályi. 2014 Applications of flow in human development and education: The collected works of Mihaly Csíkszentmihályi. Springer Science + Business Media, New York, NY, US. pp. 494, xxii, 494–xxii.

[21] Anna Karelina, Eugenia Ektina, Peter Bohacek, Matthew Vonk, Michael Kagan, Aaron R. Warren, and David T. Brookes. 2022. Comparing students' flow states during apparatus-based versus video-based lab activities. European Journal of Physics 43, 10.1088/1361-6404/ac683f.

**APPENDIX**

## A  COMPUTER SCIENCE CONCEPT REPRESENTATIONS IN EXPERIMENTAL GROUPS

| Computer Science Concept | Group A: Text Box Contents | Group B: Visual Metaphor |
|---|---|---|
| Start of program execution | Start | Green flag |
| Print statement | Print: | Printer |
| String (text) | String contents enclosed in double quotes | Open book with string contents enclosed in double quotes |
| Integer | Literal integer | 3D text of integer |
| `True` (Boolean) | True | Green traffic light |
| `False` (Boolean) | False | Red traffic light |
| Boolean variable | Name of Boolean variable (alibi) | Traffic light with no green or red light showing |
| Addition | + | 3D text of + sign |
| Multiplication | × | 3D text of × sign |
| Equality (==) | == | Balanced brass scales |
| Greater than (>) | > | Uneven brass scales |
| And (`&&`) | and | Handshake |
| Not (`!`) | not | Reverse symbol playing card |
| `if` | If: | Garden gate |

## B  QUESTIONNAIRES AND TESTS
## B.1 PRE-INTERVENTION QUESTIONNAIRE / TEST

Section 1. Tracking and Screening Information

1. What is your Subject Number?

Please ask the researcher for your subject number, if it has not been provided to you already. This is used instead of other identifying information to anonymize your responses.

Questions 2 and 3 were also on the screening survey, but responses are collected separately here, once again for the purpose of anonymity, since the screening survey also collected name and email information. The information from the screening survey will not be included in the study results.

2. When is the most recent time that you took a computer science or computer programming course?

This could include classes such as in high school or college, or online programming courses that might teach specific programming languages.

[  ] Within the past two years
[  ] More than two years ago
[  ] Never

3. When is the most recent time that you engaged in computer programming for reasons other than taking a course?

This could include reasons such as for work, a hackathon event, or teaching a computer science or computer programming course.

[  ] Within the past two years
[  ] More than two years ago
[  ] Never

Section 2. Computer Science Self-Efficacy

For the questions in this section, please answer on a 1-7 scale anywhere from [1: Not at all confident] to [7: Completely confident].

For some questions in this section, the term "block" refers to a unit of a program that may have some data as its single input or multiple inputs, and may have some data that the block outputs. Blocks may take the outputs from other blocks to use as their own inputs.

For some questions in this section, the terms "expression" and "statement" refer to pieces textual code in a written in a programming language as part of a line of code or as a full line of code.

Programming blocks are the visual programming equivalent of expressions and statements.

4. How confident are you that you can identify different basic data types used to create programs?

5. How confident are you that you can evaluate simple expressions or blocks correctly as they would be evaluated by a computer program?

6. How confident are you that you can create simple programs using either expressions and statements or programming blocks?

7. For this question, consider a simple program consisting of either 5-10 expressions and statements or 5-10 programming blocks.

If the program is not doing what it was originally intended to do, how confident are you that you can identify why that is occurring?

8. For this question, consider a simple program consisting of either 5-10 expressions and statements or 5-10 programming blocks.

If you have already identified why the program is not doing what it was intended to do, how confident are you that you can make the appropriate changes to the program so that it does what was intended?

Section 3. Computer Science Knowledge

For some questions in this section, the term "block" refers to a unit of a program that may have some data as its single input or multiple inputs, and may have some data that the block outputs. For their inputs, blocks may use the outputs from other blocks.

The term "operator block" refers to a unit of a program that definitely has some data as its single input or multiple inputs, and definitely has some data that it outputs.

If you don't know the answer to a question, you have the option to select "I don't know." You may also guess if you think you might know the answer but you aren't sure.

9. Which of the following types of data can be used when programming? You may select multiple answer responses.

[  ] numeric values
[  ] image values
[  ] text-based values
[  ] logic values
[  ] none of the above
[  ] I don't know

10. Which of the following types of data can be printed with a block that prints the input given to it?

[  ] numeric values

[  ] image values

[  ] text-based values

[  ] logic values

[  ] none of the above

[  ] I don't know

11. For this question, refer to the definition of "operator block" at the top of this section.

When can an operator block have inputs that are different types of data?

[  ] A) Always

[  ] B) Never

[  ] C) Sometimes

[  ] I don't know

12. A block compares whether two numeric values are the same. What type of data does this block output?

[  ] A) A numeric value

[  ] B) A logic value

[  ] C) A text-based value

[  ] D) None of the above

[  ] I don't know

13. For this question, consider the following blocks.

Block A: text value "Hello"

Block B:  text value  "World"

Block C:  text value  "!"

Block D puts makes a single text value out of its first and second input values by connecting them one after the other, with the first input first and the second input second.

Block E prints the input text value it is given.

-----------

If the following arrangement of blocks is used, what will be printed by Block E?

-----------

Block A and Block B are input to one Block D (called Block D1) as the first and second inputs, respectively.

The output of Block D1 and Block C are input to a second Block D (called Block D2) as the first and second inputs, respectively.

The output of Block D2 is input to Block E.

[  ] A) Hello World!
[  ] B) Hello World !
[  ] C) HelloWorld!
[  ] D) none of the above
[  ] I don't know

14. For this question, consider the following blocks.

Block A: text value "4>3"

Block B: text value " is "

Block C: outputs True if its first input is less than its second input. Otherwise it outputs False. First input: 6. Second input: 5.

Block D puts makes a single text value out of its first and second input values by connecting them one after the other, with the first input first and the second input second.

Block E prints the input text value it is given.

-----------

If the following arrangement of blocks is used, what will be printed by Block E?

-----------

Block A and Block B are input to one Block D (called Block D1) as the first and second inputs, respectively.

The output of Block D1 and Block C are input to a second Block D (called Block D2) as the first and second inputs, respectively.

The output of Block D2 is input to Block E.

[  ] A) 4>3 is True
[  ] B) 4>3 is False

[  ] C) True is (6<5)

[  ] D) False is (6<5)

[  ] I don't know

15. How many different kinds of basic logical operators are there?

[  ] A) 0

[  ] B) 2

[  ] C) 3

[  ] D) 4

[  ] I don't know

16. For this question, consider the following blocks.

Block A outputs a thumbs up if its first input is greater than its second input. Otherwise it outputs a thumbs down.

Block B has only one input. If its input is a thumbs up, it outputs a thumbs down. If its input is thumbs down, it outputs a thumbs up.

Block C outputs a thumbs up if its two inputs are the same. Otherwise it outputs a thumbs down.

Block D outputs a thumbs up if either of its two inputs is a thumbs up. Otherwise it outputs a thumbs down.

Block E outputs a thumbs up only if both of its two inputs are thumbs up. Otherwise it outputs a thumbs down.

-----------

One Block A (called A1) is given 5 as its first input and 6 as its second input.

The output of Block A1 is given as the single input to Block B.

Block C is given "cat" and "dog" as inputs.

The outputs of Block B and Block C are given as inputs to Block D.

The output of Block D is given as the first input to Block E. The second input of Block E is the output of another Block A (called A2) which has 1 for its first input and 4 for its second input.

-----------

What is the output of Block E?

[  ] A) a thumbs up
[  ] B) a thumbs down
[  ] C) both A and B
[  ] D) the output cannot be determined from the information provided
[  ] I don't know

17. For this question, consider the following blocks.

Block A outputs a green light if its first input is equal to its second input. Otherwise it outputs a red light. It is given 1 as its first input and 2 as its second input.

Block B prints "EXCLAMATION" if its input is a green light. Otherwise it will do nothing besides move on to the next block if there is one.

Block C prints "!" (If there is any text already printed, Block C prints "!" on the same line.)

-----------

The output of Block A is the input for Block B. There is a sequence made up of Block B followed by Block C.

-----------

What will be printed by the sequence?

[  ] A) EXCLAMATION
[  ] B) !
[  ] C) EXCLAMATION!
[  ] D) None of the above
[  ] I don't know

18. For this question, refer to the definitions for "block" and "operator block" at the top of this section.

Complete the following sentence:

One feature that coding blocks like printing blocks and operator blocks have in common is that they must have ___________.

[  ] A) one or more data inputs
[  ] B) data outputs that can be used by other coding blocks as inputs
[  ] C) coding blocks that happen before them
[  ] D) coding blocks that happen after them
[  ] I don't know

## B.2 POST-INTERVENTION QUESTIONNAIRE / TEST

Section 1. Subject Number

1. What is your Subject Number?

Please ask the researcher for your subject number, if it has not been provided to you already. This is used instead of other identifying information to anonymize your responses.

Section 2. VR Activity Engagement

2. For the VR activity, to what extent was assistance needed from the researcher?

Please answer on a 1-7 scale anywhere from [1: Not at all] to [7: A lot]

3. For the VR activity, to what extent did you know what to do (goal of the tasks)?

Please answer on a 1-7 scale anywhere from [1: Not at all] to [7: A lot]

4. For the VR activity, to what extent did you know how to do it (the goal of the tasks)?

Please answer on a 1-7 scale anywhere from [1: No idea] to [7: Completely]

5. For the VR activity, to what extent did you know how well you were doing?

Please answer on a 1-7 scale anywhere from [1: No idea] to [7: Completely]

6. To what extent was the VR activity challenging?

Please answer on a 1-7 scale anywhere from [1: Not at all] to [7: Extremely]

7. To what extent did you feel knowledgeable and skillful during the VR activity?

Please answer on a 1-7 scale anywhere from [1: Not at all] to [7: Extremely]

8. To what extent was the VR activity fun and interesting?

Please answer on a 1-7 scale anywhere from [1: Not at all] to [7: Extremely]

Section 3. Computer Science Self-Efficacy

For the questions in this section, please answer on a 1-7 scale anywhere from [1: Not at all confident] to [7: Completely confident].

For some questions in this section, the term "block" refers to a unit of a program that may have some data as its single input or multiple inputs, and may have some data that the block outputs. Blocks may take the outputs from other blocks to use as their own inputs.

For some questions in this section, the terms "expression" and "statement" refer to pieces textual code in a written in a programming language as part of a line of code or as a full line of code.

Programming blocks are the visual programming equivalent of expressions and statements.

9. How confident are you that you can identify different basic data types used to create programs?

10. How confident are you that you can evaluate simple expressions or blocks correctly as they would be evaluated by a computer program?

11. How confident are you that you can create simple programs using either expressions and statements or programming blocks?

12. For this question, consider a simple program consisting of either 5-10 expressions and statements or 5-10 programming blocks.

If the program is not doing what it was originally intended to do, how confident are you that you can identify why that is occurring?

13. For this question, consider a simple program consisting of either 5-10 expressions and statements or 5-10 programming blocks.

If you have already identified why the program is not doing what it was intended to do, how confident are you that you can make the appropriate changes to the program so that it does what was intended?

Section 4. Computer Science Knowledge

For some questions in this section, the term "block" refers to a unit of a program that may have some data as its single input or multiple inputs, and may have some data that the block outputs. For their inputs, blocks may use the outputs from other blocks.

The term "operator block" refers to a unit of a program that definitely has some data as its single input or multiple inputs, and definitely has some data that it outputs.

If you don't know the answer to a question, you have the option to select "I don't know." You may also guess if you think you might know the answer but you aren't sure.

14. Which of the following types of data can be used when programming? You may select multiple answer responses.

[  ] numeric values

[  ] image values

[  ] text-based values

[  ] logic values

[  ] none of the above

[  ] I don't know

15. Which of the following types of data can be printed with a block that prints the input given to it?

[  ] numeric values

[  ] image values

[  ] text-based values

[  ] logic values

[  ] none of the above

[  ] I don't know

16. For this question, refer to the definition of "operator block" at the top of this section.

When can an operator block have inputs that are different types of data?

[  ] A) Always

[  ] B) Never

[  ] C) Sometimes

[  ] I don't know

17.  A block compares whether if one numeric value is greater than a second numeric value? What type of data does this block output?

[  ] A) A numeric value

[  ] B) A logic value

[  ] C) A text-based value

[  ] D) None of the above

[  ] I don't know

18. For this question, consider the following blocks.

Block A: text value  "How"

Block B:  text value  "are you"

Block C:  text value  "?"

Block D puts makes a single text value out of its first and second input values by connecting them one after the other, with the first input first and the second input second.

Block E prints the input text value it is given.

-----------

If the following arrangement of blocks is used, what will be printed by Block E?

-----------

Block A and Block B are input to one Block D (called Block D1) as the first and second inputs, respectively.

The output of Block D1 and Block C are input to a second Block D (called Block D2) as the first and second inputs, respectively.

The output of Block D2 is input to Block E.

[  ] A) How are you?

[  ] B) How are you ?

[  ] C) Howare you?

[  ] D) none of the above

[  ] I don't know

19. For this question, consider the following blocks.

Block A: text value "1+2=3"

Block B: text value " is "

Block C: outputs True if its first input is greater than its second input. Otherwise it outputs False. First input: 7. Second input: 4.

Block D puts makes a single text value out of its first and second input values by connecting them one after the other, with the first input first and the second input second.

Block E prints the input text value it is given.

-----------

If the following arrangement of blocks is used, what will be printed by Block E?

-----------

Block A and Block B are input to one Block D (called Block D1) as the first and second inputs, respectively.

The output of Block D1 and Block C are input to a second Block D (called Block D2) as the first and second inputs, respectively.

The output of Block D2 is input to Block E.

[  ] A) 1+2=3 is True

[  ] B) 1+2=3 is False

[  ] C) True is (7>4)

[  ] D) False is (7>4)

[  ] I don't know

20. How many different kinds of basic logical operators are there?

[  ] A) 0

[  ] B) 2

[  ] C) 3

[  ] D) 4

[  ] I don't know

21. For this question, consider the following blocks.

Block A outputs a smiley face if its first input is the same as its second input. Otherwise it outputs a frowny face.

Block B has only one input. If its input is a smiley face, it outputs a frowny face. If its input is a frowny face, it outputs a smiley face.

Block C outputs a smiley face if its first input is less than its second input. Otherwise it outputs a frowny face.

Block D outputs a smiley face only if both of its two inputs are smiley faces. Otherwise it outputs a frowny face.

Block E outputs a smiley face if either of its two inputs is a smiley face. Otherwise it outputs a frowny face.

-----------

One Block A (called A1) is given "bird" as its first input and "bird" as its second input.

The output of Block A1 is given as the single input to Block B.

Block C is given 5 as its first input and 1 as its second input.

The outputs of Block B and Block C are given as inputs to Block D.

The output of Block D is given as the first input to Block E. The second input of Block E is the output of another Block A (called A2) which has 6 for its first input and 8 for its second input.

-----------

What is the output of Block E?

[  ] A) a smiley face
[  ] B) a frowny face
[  ] C) both A and B
[  ] D) the output cannot be determined from the information provided
[  ] I don't know

22. For this question, consider the following blocks.

Block A outputs a star if its first input is equal to its second input. Otherwise it outputs an X. It is given 7 as its first input and 7 as it second input.

Block B prints "WOOHOO" if its input is a star. Otherwise it will do nothing besides move on to the next block if there is one.

Block C prints "..." (If there is any text already printed, Block C prints "..." on the same line.)

-----------

The output of Block A is the input for Block B. There is a sequence made up of Block B followed by Block C.

-----------

What will be printed by the sequence?

[  ] A) WOOHOO
[  ] B) ...
[  ] C) WOOHOO...
[  ] D) None of the above
[  ] I don't know